\documentstyle[12pt,fleqn]{article}
\topmargin - 0.8cm

\catcode`@=11
\@addtoreset{equation}{section}
\catcode`@=12
\mathchardef\SGamma="7100

\begin{document}
\title{\bf Open inflation from quantum cosmology with a strong nonminimal coupling}
\author{A.O. Barvinsky$^{\dag}$}
\date{}
\maketitle
\hspace{-8mm}{\em
Theory Department, Lebedev Physics Institute and Lebedev Research Center in
Physics, Leninsky Prospect 53,
Moscow 117924, Russia}
\begin{abstract}
We propose the mechanism of quantum creation of the open Universe in the
observable range of values of $\Omega$. This mechanism is based on the
no-boundary quantum state with the Hawking-Turok instanton applied to
the model with a strong nonminimal coupling of the inflaton field. We
develop the slow roll perturbation expansion for the instanton solution
and obtain a nontrivial contribution to the classical instanton action.
The interplay of this classical contribution with the loop effects due to
quantum effective action generates the probability distribution peak with
necessary parameters of the inflation stage without invoking any anthropic
considerations. In contrast with a similar mechanism for closed models,
existing only for the tunneling quantum state of the Universe, the
observationally justified open inflation originates from the no-boundary
cosmological wavefunction.
\end{abstract}
$^{\dag}$e-mail: barvin@td.lpi.ac.ru\\
PACS: 98.80.Hw, 98.80.Bp, 04.60.-m\\
Keywords: Quantum cosmology; inflation; effective action

\section{Introduction}
\hspace{\parindent}
Hawking and Turok have recently suggested the mechanism of quantum
creation of an open Universe from the no-boundary cosmological state
\cite{HawkTur}. Motivated by the observational evidence for inflationary
models with $\Omega<1$ they constructed a singular gravitational
instanton capable of generating expanding universes with open spatially
homogeneous sections.
The prior quantum probability of such universes weighted by the anthropic
probability of galaxy formation was shown to be peaked at $\Omega\sim 0.01$.
This idea, despite its extremely attractive nature, was criticized from
various sides. In order to increase the amount of inflation to larger values
of $\Omega$ and avoid anthropic considerations Linde \cite{Lindop} proposed
to replace the no-boundary quantum state \cite{HH,H} by the tunneling one
\cite{VilNB,tun}. The singularity
of the Hawking-Turok instanton raised a number of objections both in the
Euclidean theory \cite{BousLind,Vilen,Unruh,Wu,Garriga} and from the
viewpoint of the resulting timelike singularity in the expanding Universe
\cite{Unruh,Starobin,Garriga}. The criticism of singular instantons was
followed by attempts of their justification \cite{HawkTur1,Garriga1} which
still leave their issue open.

In any case it seems that the practical goal of quantum cosmology --
generating the open Universe with observationally justified modern value of
$\Omega$, not very close to one or zero, -- has not yet been reached. The use
of anthropic principle, as was recognized by the authors of \cite{HawkTur},
is certainly a retreat in theory, because by and large this principle has
such a disadvantage that it can explain practically everything without
being able to predict anything. The tunneling state advocated by Linde
\cite{Lindop} (and strongly criticized in \cite{HawkTur2}) requires special
supergravity induced potentials and takes place at energies beyond reliable
perturbative domain with the resulting $\Omega\simeq 1$. Other works in the
above series discuss conceptual issues of the Hawking-Turok proposal
without offering the concrete mechanism of generating the needed $\Omega$.

On the other hand, in spatially {\it closed} context there exists a
mechanism of generating the probability peak in the distribution of
cosmological models at a low (typically GUT) energy scale. It does not
appeal to anthropic considerations. Rather it is based on quantum loop
effects \cite{norm,tunnel,BarvU} in the model of chaotic inflation with large
negative nonminimal coupling of the inflaton \cite{qsi,qcr,tvsnb,efeq}. This peak
exists for a wide class of particle physics models of GUT type, coupled to
the inflaton. Because of its GUT energy scale its mechanism is not
essentially affected by the nonrenormalizability problems inherent to
quantum gravity, or by modifications due to a more fundamental theory like
superstrings. Moreover, big negative nonminimal coupling of the inflaton
plays in this model the role of the effective inverse gravitational
constant, which improves the loop expansion and suppresses the
contributions of higher spins including gravitons. Thus it seems to be
robust against nonperturbative issues. Important peculiarity of this
result is that it exists only for the tunneling cosmological wavefunction,
because a similar peak for the no-boundary state of the Universe, existing
in certain domain of coupling constants, can generate only an infinitely
long inflationary stage.

In this paper we extend this mechanism to the {\it open} Universe and
show, what was briefly reported in recent author's paper \cite{noanthr},
that the situation qualitatively reverses: the probability peak in
the distribution of open inflationary models exists for
the no-boundary state based on the Hawking-Turok instanton. Similarly to
\cite{qsi,tvsnb,qcr,efeq} it has GUT scale parameters and the value of $N$ easily
adjustable (without fine tuning of initial conditions and anthropic
considerations) for observationally justified values of $\Omega$. We begin
in Sect.2 with a brief overview of the closed model \cite{qsi,tvsnb,qcr,efeq}
with a special emphasis on the interplay between the classical Euclidean
action and its quantum effective counterpart, generating the probability
peak. Sect.3 is devoted to the construction of the Hawking-Turok instanton
and a special algorithm for its classical action, based on the asymptotic
behaviour of fields near the singularity. In Sect.4 this construction is
extended to the model with a nonminimal inflaton field. Sects.5 and 6
contain a detailed presentation of the slow roll perturbation theory for
the Hawking-Turok solution of field equations and the calculation of its
action. Remarkably, this tree-level action has a large positive
contribution logarithmic in the inflaton field. It is structurally analogous to
loop corrections and already at the tree level is capable of generating the
inflation probability peaks. In Sect.7 we show that these peaks do not
satisfy the observational constraints, because they generate either too low,
$\Omega\sim e^{-110}$, or too close to one, $\Omega\sim 1-e^{-2\times
10^6}$, values of the density parameter. For this reason, in Sect.8 we turn
to quantum loop effects on the Hawking-Turok instanton. We discuss the
problems with the quantum effective action associated with its singularity
and find that the dominant scaling behaviour is robust against this
singularity. Finally, in Sect.9 we calculate the most probable $H$, $N$
and $\Omega$ of the open inflation generated within the no-boundary
Hawking-Turok paradigm. In the concluding Sect.10 we discuss the modifications
of the obtained results by loop corrections to the classical equations of
motion \cite{efeq} at the stage of the inflationary evolution and
show that they do not qualitatively change the conclusions.

\section{Quantum cosmological origin of the closed inflation}
\hspace{\parindent}
Quantum cosmology serves as a theory of quantum initial conditions for the
evolution of our Universe. One of the main observable cosmological parameters
is the density parameter $\Omega$, and in the model, undergoing the
conventional inflationary and standard big-bang stages, its origin can be
traced back to the initial conditions for inflation. Matching the
inflationary, radiation and then matter dominated stages leads to the
following expression for the present day value of $\Omega$ \cite{HawkTur},
        \begin{eqnarray}
        \Omega\simeq\frac1{1\mp B\exp(-2N)},        \label{i1}
        \end{eqnarray}
in terms of the inflationary e-folding number $N$ -- the logarithmic
expansion coefficient for the cosmological scale factor $a$ during the
inflation stage with a Hubble constant $H=\dot a/a$,
        \begin{eqnarray}
        N=\int_0^{t_F}dt H                     \label{i2}
        \end{eqnarray}
(with $t=0$ and $t_F$ denoting the beginning and the end of inflation
epoch). The signs $\mp$ in (\ref{i1}) are related respectively to the closed
and open models, and $B$ is the parameter incorporating the details of the
reheating and radiation-to-matter transitions in the early Universe.
Depending on the model for these transitions, its order of magnitude can
vary from $10^{25}$ to $10^{50}$ (when the reheating temperature varies
from the electroweak to GUT scale). In what follows we shall assume the
latter as the most probable value of this parameter.

Eq. (\ref{i1}) clearly demonstrates rather stringent bounds on $N$. For
the closed model  the e-folding number should satisfy the lower bound
$N\geq 60$ in order to generate the observable Universe of its present
size, while for the open model $N$ should be very close to this bound
$N\simeq 60$ in order to have the present value of $\Omega$ {\em not very
close to zero or one}, $0<\Omega<1$, the fact intensively discussed on the
ground of the recent observational data.

In the chaotic inflation model the effective Hubble constant is generated
by the potential of the inflaton scalar field and all the parameters of
the inflationary epoch, including its duration in units of $N$, can be
found as functions of the initial value of the inflaton field $\varphi$
at the onset of inflation $t=0$. This initial condition belongs to the
quantum domain, i.e. it has to be considered subject to the quantum
distribution following from the cosmological wavefunction. If this
distribution function has a sharp probability peak at certain $\varphi$,
then, at least within the semiclassical expansion, this value of $\varphi$
serves as the initial condition for the inflationary dynamics.

There are two known quantum states that lead in the semiclassical
regime to the closed chaotic inflation Universe. They are given by the
no-boundary \cite{HH,H,VilNB} and tunneling \cite{tun} wavefunctions.
In the tree-level approximation they generate the distribution functions
which are unnormalizable in the high-energy domain $\varphi\to\infty$ and
generally devoid of the observationally justfied probability peaks. It
turned out, however, that the inclusion of quantum loop effects can
qualitatively change the situation -- make the distributions normalizable
\cite{norm,BarvU}.  Moreover, in the closed model with a strong nonminimal
coupling of the inflaton to curvature, these effects can generate the
probabilty peak at GUT energy scale satisfying the above bound on $N$
\cite{qsi,tvsnb,qcr}. The basic formalism underlying this result is as
follows.

In the quantum gravitational domain the
conventional expression for the no-boundary and tunneling distributions of
the inflaton field $\rho_{\rm NB,T}(\varphi)\sim\exp[\mp I(\varphi)]$ is
replaced by
        \begin{eqnarray}
        \rho_{\rm NB,T}(\varphi)\sim\exp[\mp I(\varphi)-
        \mbox{\boldmath$\SGamma$}(\varphi)],          \label{0.1}
        \end{eqnarray}
where the classical Euclidean action $I(\varphi)$ on the quasi-DeSitter
instanton with
the inflaton value $\varphi$ is amended by the loop effective action
$\mbox{\boldmath$\SGamma$}(\varphi)$ calculated on the same instanton
\cite{norm,tunnel,BarvU,tvsnb}. The contribution of the latter can qualitatively
change predictions of the tree-level theory due to the dominant part of the
effective action induced by the anomalous scaling behaviour. On the instanton
of the size $1/H(\varphi)$ -- the inverse of the Hubble constant, it looks like
        \begin{eqnarray}
        \mbox{\boldmath$\SGamma$}(\varphi)\sim Z\ln H(\varphi),
        \end{eqnarray}
where $Z$ is the total anomalous scaling of all quantum fields in the model.
For the model of \cite{qsi,qcr}
        \begin{equation}
        {\mbox{\boldmath $L$}}(g_{\mu\nu},\varphi)
        =g^{1/2}\left\{\frac{m_{P}^{2}}{16\pi} R(g_{\mu\nu})
        -\frac{1}{2}\xi\varphi^{2}R(g_{\mu\nu})
        -\frac{1}{2}(\nabla\varphi)^{2}
        -\frac{1}{2}m^{2}\varphi^{2}
        -\frac{\lambda}{4}\varphi^{4}\right\},      \label{0.2}
        \end{equation}
with a big negative constant $-\xi=|\xi|\gg 1$ of nonminimal curvature
coupling, and generic GUT sector of Higgs $\chi$, vector gauge $A_\mu$
and spinor fields $\psi$ coupled to the inflaton via the interaction term
        \begin{eqnarray}
        {\mbox{\boldmath $L$}}_{\rm int}
        =\sum_{\chi}\frac{\lambda_{\chi}}4
        \chi^2\varphi^2
        +\sum_{A}\frac12 g_{A}^2A_{\mu}^2\varphi^2+
        \sum_{\psi}f_{\psi}\varphi\bar\psi\psi
        +{\rm derivative\,\,coupling},             \label{0.3}
        \end{eqnarray}
this parameter can be very big, because of the Higgs effect generating large
masses of all the corresponding particles directly coupled the inflaton. Due to
this effect the parameter $Z$ (dominated by terms quartic in particle masses)
is quadratic in $|\xi|$,
        \begin{eqnarray}
        &&Z=6\frac{|\xi|^2}\lambda\mbox{\boldmath$A$},    \label{Z}\\
        &&{\mbox{\boldmath $A$}} = \frac{1}{2\lambda}
        \Big(\sum_{\chi} \lambda_{\chi}^{2}
        + 16 \sum_{A} g_{A}^{4} - 16
        \sum_{\psi} f_{\psi}^{4}\Big),               \label{A}
        \end{eqnarray}
with a universal combination of the coupling constants above.

Thus, the probability peak in this model reduces to the extremum of
the function
        \begin{eqnarray}
        \ln\rho_{\rm NB,\,T}(\varphi)\simeq\mp I(\varphi)
        -3\frac{|\xi|^2}\lambda\mbox{\boldmath$A$}\,
        \ln\frac{\varphi^2}{\mu^2}.                    \label{1.1}
        \end{eqnarray}
in which the $\varphi$-dependent part of the classical instanton action
        \begin{eqnarray}
        &&I(\varphi)=-\frac{96\pi^2|\xi|^2}\lambda
        -\frac{24\pi(1+\delta)|\xi|}
        {\lambda}\frac{m_P^2}{\varphi_0^2}
        +O\,\left(\frac{m_P^4}{\varphi^4}\right),     \label{1.2} \\
        &&\delta\equiv
        -\frac{8\pi\,|\xi|\,m^2}{\lambda\,m_P^2},    \label{delta}
        \end{eqnarray}
should be balanced by the anomalous scaling term, provided the signs
of $(1+\delta)$ and $\mbox{\boldmath$A$}$ are properly correlated with
the $(\mp)$ signs of the no-boundary (tunneling) proposals. As a result
the probability peak exists with parameters -- mean values of the inflaton
and Hubble constants and relative width
        \begin{eqnarray}
        &&\varphi_I^2= m_{P}^2\frac{8\pi|1+\delta|}{|\xi|
        {\mbox{\boldmath$A$}}},             \label{1.3}\\
        &&H^2(\varphi_I)=
        m_{P}^2\frac{\lambda}{|\xi|^2}
        \frac{2\pi|1+\delta|}
        {3{\mbox{\boldmath $A$}}},                   \label{1.3a}\\
        &&\frac{\Delta\varphi}{\varphi_I}\sim
        \frac{\Delta H}{H}\sim
        \frac 1{\sqrt{12{\mbox{\boldmath $A$}}}}
        \frac{\sqrt{\lambda}}{|\xi|},              \label{1.4}
        \end{eqnarray}
which are strongly suppresed by a small ratio $\sqrt{\lambda}/|\xi|$
known from the COBE normalization for $\Delta T/T\sim 10^{-5}$
\cite{COBE,Relict}(because
the CMBR anisotropy in this model is proportional to this ratio
\cite{nonmin}). This GUT scale peak gives rise to the finite
inflationary epoch with the e-folding number
        \begin{eqnarray}
        N\simeq \frac{48\pi^2}{\mbox{\boldmath $A$}}.   \label{1.5}
        \end{eqnarray}
only for $1+\delta>0$ and, therefore, only for the {\it tunneling} quantum
state (plus sign in (\ref{1.1})). Comparison with $N\geq 60$ necessary
for $\Omega>1$ immeadiately yields the bound on
$\mbox{\boldmath $A$}$ \cite{efeq},
        \begin{eqnarray}
        \mbox{\boldmath $A$}\sim 5.5,               \label{bound1}
        \end{eqnarray}
which can be regarded as a
selection criterion for particle physics models \cite{qsi}. This conclusions
on the nature of the inflation dynamics from the initial probability peak
remain true also at the quantum level -- with the effective equations
replacing the classical equations of motion \cite{efeq}.

For the proponents of the no-boundary quantum states in a long debate on the
wavefunction discord \cite{VilVach,HawkTur2,VilLorentz,discorde} this
situation might seem unacceptable. According to this result the
no-boundary proposal does not generate realistic inflationary scenario,
while the tunneling state does not satisfy important aesthetic criterion
-- the universal formulation of the initial conditions and dynamical
aspects in one concept -- spacetime covariant path integral over
geometries\footnote{The Lorentzian path integral for the tunneling state
of \cite{VilLorentz} also requires, in this respect, extension beyond
minisuperspace level, development of the semiclassical expansion technique,
etc.}. In what follows we show that for the open model this situation
qualitatively reverses.

\section{The Hawking-Turok instanton in a minimal model}
\hspace{\parindent}
We assume that the reader is familiar with the construction of the
Hawking-Turok instanton generating the open inflation. Very briefly, the
inflating Lorentzian spacetime originates in both closed and open models
by the nucleation from a 3-dimensional section of the gravitational
instanton. In the closed model this is the equatorial section -- the
boundary of the 4-dimensional quasi-hemishpere labelled by the constant value
of the latitude anglular coordinate. The analytic continuation of this
coordinate into the complex plane gives rise to the Lorentzian
quasi-DeSitter spacetime modelling the open inflation. In the open case
the Hawking-Turok suggestion was to continue the Euclidean solution beyond
the equatorial section up to the point where the Euclidean scale factor
again vanishes at the point antipoidal to the regular pole on the first
hemisphere. The nucleation surface then has to be chosen as the
longitudinal section of this quasisherical manifold passing through the
regular pole and its antipoidal point. Then the analytic continuation of
the corresponding longitudinal angle into the complex plane gives rise to
the Lorentzian spacetime. The light cone originating from the regular pole
cuts in this spacetime the domain sliced by the open spatially homogeneous
sections of constant negative curvature. Their chronological succession
serves as the model for the open inflationary Universe.

The difficulty
with this construction follows from the fact that the point antipoidal to
the regular pole of the Hawking-Turok instanton is not regular -- it has a
singularity which develops into the time-like singularity in the
Lorentzian part of spacetime. The main objections against the
Hawking-Turok instanton are targeting this singularity. We shall not,
however, dwell on the criticism of this construction, and adopt the
treatment of this singularity suggested in \cite{HawkTur1}. It should be
regarded as point punctured from the manifold by choosing a surrounding it
boundary surface shrinking to the point as a limiting procedure. The usual
Dirichlet boundary conditions for boundary metric coefficients are assumed
at this surface, so that the Eistein Euclidean action on the Hawking-Turok
instanton has the form \cite{HawkTur}
        \begin{eqnarray}
        I[g_{\mu\nu},\phi]=\int_{\cal M} d^4x\, g^{1/2}\left\{
        -\frac{m_P^2}{16\pi}R+\frac12(\nabla\phi)^2+V(\phi)\right\}
        +\frac{m_P^2}{8\pi}
        \int_{\partial\cal M}d^3x\, ({}^3g)^{1/2} K,    \label{2.1}
        \end{eqnarray}
with $K$ -- the extrinsic curvature at this boundary.  Here we consider
the model with minimally coupled inflaton field denoted by $\phi$ (in
contrast with $\varphi$ reserved for the nonminimally coupled inflaton)
and having the inflaton potential $V(\phi)$.

With the spatially homogeneous ansatz for the Euclidean metric
($d\Omega^2_{(3)}$ is the metric of the 3-dimensional sphere
of unit radius, $a(\sigma)$ is the scale factor and $\sigma$ is the
latitude angular coordinate of the above type),
        \begin{eqnarray}
        ds^2=d\sigma^2+a^2(\sigma)\,d\Omega^2_{(3)} ,    \label{2.2}
        \end{eqnarray}
the Euclidean equations of motion take the form
        \begin{eqnarray}
        &&\ddot\phi+3\frac{\dot a}a\dot\phi-V'=0,  \label{2.3}\\
        &&a^3V-\frac{3m_P^2}{8\pi}(a-a\dot a^2)
        -\frac12 a^3\dot\phi^2=0,                      \label{2.4}
        \end{eqnarray}
where dots denote the derivatives with respect to the coordinate $\sigma$
and the prime denotes the derivative with respect to the inflaton scalar
field.

The regularity of the instanton solution at $\sigma=0$ implies that
$\dot\phi(0)=0$ and $\dot a(0)=1$ (the absence of the conical singularity).
With these initial conditions and when the slope of the inflaton potential
is not too steep the solution near $\sigma=0$ can be obtained in the slow roll
approximation as an expansion in powers of the dimensionless parameter
        \begin{eqnarray}
        \epsilon=\frac{m_P}{\sqrt{3\pi}}
        \frac{V'(\phi_0)}{V(\phi_0)}.                     \label{2.7}
        \end{eqnarray}
In the lowest order approximation this solution represents the constant
inflaton field $\phi_0=\phi(0)$ and the scale factor of the Euclidean DeSitter
geometry
        \begin{eqnarray}
        &&a=\frac1{H_0} \sin\theta+\delta a,\,\,\,
        \phi=\phi_0+\delta\phi,                \label{2.5} \\
        &&H_0^2=\frac{8\pi V(\phi_0)}{3m_P^2},\,\,\,
        \theta=H_0\sigma,                            \label{2.6}\\
        &&\delta\phi(\theta)=m_P\,O(\epsilon),\,\,\
        \delta a(\theta)=\frac1{H_0}\,
        O(\epsilon^2),                  \label{2.9}
        \end{eqnarray}
with the effective Hubble constant $H_0$ given in terms of the initial
value of the potential $V(\phi_0)$.

The slow roll approximation does not, however, hold for all values of
$\sigma$. As it follows from the dynamical equation for $a$,
        \begin{eqnarray}
        \ddot a+\frac{8\pi}{3m_P^2}(V+\dot\phi^2)a=0,
        \end{eqnarray}
for monotonically growing positive potentials the scale
factor has a negative second derivative which converts at certain moment
its expansion to a contraction and then makes it vanishing at some
$\sigma=\sigma_f$, $a(\sigma_f)=0$. Near this point, kinetic terms of the
equations above start dominating and, thus, allow one to write the
approximate solution \cite{HawkTur}
        \begin{eqnarray}
        &&a\simeq A(\sigma_f-\sigma)^{1/3},        \label{2.10a}\\
        &&\phi\simeq-\frac{m_P}{\sqrt{12\pi}}
        \ln(\sigma_f-\sigma)+\Phi,\,\,\,
        \sigma\rightarrow\sigma_f.                 \label{2.10}
        \end{eqnarray}
It is important that the coefficient of the logarithmic singularity of the
scalar field is unambiguously defined from the equations of motion, whereas
the coefficients $A$ and $\Phi$ nontrivially depend on the initial condition
at $\sigma=0$, that is on $\phi_0$.

The knowledge of $A(\phi_0)$ and $\Phi(\phi_0)$ allows one to
obtain the action of the Hawking-Turok instanton. Its action (\ref{2.1})
for the metric (ref{2.2}) has the
        \begin{eqnarray}
        I_{HT}(\phi_0)=2\pi^2\int_0^{\sigma_f}d\sigma\,\left[a^3V
        -\frac{3m_P^2}{8\pi}a(1+\dot a^2)
        +\frac12 a^3\dot\phi^2\right]
        _{\phi(\sigma,\phi_0),\,\,
        a(\sigma,\phi_0)},                           \label{2.13}
        \end{eqnarray}
where the surface term was removed by integration by parts (which
simultaneously removes the term with the second derivative of the scale
factor). It depends on $\phi_0$, and the mechanism of this dependence
originates from the behaviour of fields at the singularity. Indeed,
differentiating (\ref{2.13}) with respect to
$\phi_0$ one finds that the volume part vanishes in virtue of equations
of motion, while integrations by parts give a typical surface term involving
the Lagrangian of (\ref{2.13}) and its derivatives with respect to
$(\dot\phi,\dot a)$, which does not vanish at $\sigma_f$.

For $\sigma$ close to $\sigma_f$ the functional dependence of the fields
in (\ref{2.10}) on $\phi_0$ enters through the coefficients $A,\,\Phi$ as
well as through $\sigma_f$. Since $\sigma_f$ enters the fields in the
combination $\sigma-\sigma_f$, the total derivative of the field takes
the form
        \begin{eqnarray}
        \frac {d\phi}{d\phi_0}=
        \frac{\partial\phi}{\partial\phi_0}
        -\dot\phi\frac{\partial\sigma_f}{\partial\phi_0},  \nonumber
        \end{eqnarray}
where partial derivative with respect to $\phi_0$ acts only on $\Phi$
(and coefficients of higher powers in $(\sigma-\sigma_f)$). Similar
relation holds also for the scale factor. Thus, the surface term at
$\sigma_f$ equals
        \begin{eqnarray}
        &&\frac{dI_{HT}(\phi_0)}{d\phi_0}=\left.\left(L
        -\dot\phi\frac{\partial L}{\partial\dot\phi}
        -\dot a\frac{\partial L}{\partial\dot a}\right)
        \frac{\partial\sigma_f}{\partial\phi_0}\,
        \right|_{\,\sigma_f}\nonumber\\
        &&\qquad\qquad\qquad
        +\left(\,\frac{\partial L}{\partial\dot a}
        \frac{\partial A}{\partial\phi_0}
        (\sigma_f-\sigma)^{1/3}
        +\frac{\partial L}{\partial\dot\phi}
        \frac{\partial\Phi}{\partial\phi_0}\right)
        _{\,\sigma\rightarrow\sigma_f},                  \label{2.14}
        \end{eqnarray}
where $L$ is the Lagrangian of the Euclidean action (\ref{2.13}). The first
term here identically vanishes, because it is proportional to the Hamiltonian
constraint (in terms of velocities). On using the asymptotic behaviour near the
singularity (\ref{2.10}) one then finds
        \begin{eqnarray}
        \frac{dI_{HT}(\phi_0)}{d\phi_0}=
        \frac{\pi m_P^2}6\frac{dA^3}{d\phi_0}
        +\frac{\pi^{3/2}m_P A^3}{\sqrt 3}\frac{d\Phi}{d\phi_0},  \label{2.15a}
        \end{eqnarray}
so that the integration of this equation gives $I_{HT}(\phi_0)$.

To find $(A(\phi_0),\,\Phi(\phi_0))$ we shall develop in Sect.4 the perturbation
expansion of the solution near
$\sigma_f$. In contrast with the slow roll expansion near $\sigma=0$ this is
the expansion in powers of the potential $V(\phi)$ itself rather than
its gradient, and $\dot\phi$-derivatives give the dominant contribution
at this asymptotics. Then we match the both asymptotic expansions in the
domain of $\sigma$ where they are both valid. From this match one easily
finds the unknown parameters $A,\Phi$ as functions of
$\phi_0$. As we shall show in Sect.4, in the lowest order of the slow roll
expansion they are given by
        \begin{eqnarray}
        &&A=\left(\frac{3\epsilon}
        {H_0^2}\right)^{1/3}+O(\epsilon),                  \label{2.12a}\\
        &&\Phi=\phi_0-\frac12\frac{m_P}{\sqrt{12\pi}}
        \ln\left[\frac{9H_0^2}{8\epsilon}\right]+O(\epsilon)
        =\phi_0+\frac12\frac{m_P}{\sqrt{12\pi}}
        \ln\frac{V_0'}{V_0^2}+{\rm const}+O(\epsilon).     \label{2.12}
        \end{eqnarray}
A remarkable property of these simple expressions is that, when
substituted into (\ref{2.15a}), they yield the equation
        \begin{eqnarray}
        \frac{dI_{HT}(\phi_0)}{d\phi_0}\simeq\left(\frac d{d\phi_0}+
        \sqrt{\frac3{16\pi}}m_P\frac{d^2}
        {d\phi_0^2}\right)\,
        \left(-\frac{3m_P^4}{8V(\phi_0)}\right)+O(\epsilon),      \label{2.15}
        \end{eqnarray}
that can be integrated for a generic inflaton
potential in a closed form. In the first order of the slow roll expansion
inclusive, it reads
        \begin{eqnarray}
        I_{HT}(\phi_0)\simeq\left(1+\sqrt{\frac3{16\pi}}m_P
        \frac d{d\phi_0}\right)\,
        \left(-\frac{3m_P^4}{8V(\phi_0)}\right)+O(\epsilon^2).    \label{2.16}
        \end{eqnarray}
Note that the integration just removed one overall derivative, the
integration constant following from the known answer for $\epsilon=0$ --
the case of an exact DeSitter solution. One can check that the second term,
which represents the first order correction $O(\epsilon)$, corresponds
to the contribution of the extrinsic curvature surface part of the action
(\ref{2.1}). This expression reproduces the result of ref. \cite{HawkTur1}
obtained by indirect and less rigorous method\footnote {The calculations of
\cite{HawkTur1} do not take into account the slow roll corrections to the
volume part of the action.}. It is interesting that this algorithm is universal
for a wide class of inflaton potentials $V(\phi_0)$ ($V(\phi)$ should only
satisfy typical restrictions imposed by the slow roll approximation) and in a
closed form expresses the result in terms of $V(\phi)$ and its derivative.

Let us now go over to the model with a big nonminimal inflaton coupling
and show that the above algorithm is not sufficient in the relevant
first order of the slow roll expansion.

\section{Nonminimal model}
\hspace{\parindent}
In the model (\ref{0.2}) the inflaton field $\varphi$ is nonminimally
coupled to curvature via the effective $\phi$-dependent Planck ``mass''
$16\pi U(\phi)$
        \begin{eqnarray}
        I=\int_{\cal M} d^4x\, g^{1/2}\left\{V(\varphi)
        -U(\varphi)R+\frac12(\nabla\varphi)^2\right\}
        +2\int_{\partial\cal M}d^3x\,
        ({}^3g)^{1/2} U(\varphi)K.                    \label{3.1}
        \end{eqnarray}
For completeness we supplied this action with a boundary term necessary in
the vicinity of the singularity and also introduced the notation for the
inflaton potential $V(\varphi)$.
It is well known that this action can be reparametrized to the Einstein frame
by special conformal transformation and reparametrization of the
inflaton field $(g_{\mu\nu},\phi)\rightarrow(G_{\mu\nu},\phi)$. These
transformations are implicitly given by equations \cite{renorm}
        \begin{eqnarray}
        &&G_{\mu\nu}=\frac{16\pi U(\varphi)}
        {m_P^2}g_{\mu\nu},                            \label{3.2}\\
        &&\left(\frac{d\phi}{d\varphi}\right)^2
        =\frac{m_P^2}{16\pi}\frac{U+3U'^2}{U^2}.      \label{3.3}
        \end{eqnarray}
The action in terms of new fields
        \begin{eqnarray}
        \bar{I}=\int_{\cal M} d^4x\, G^{1/2}\left\{\bar{V}(\phi)
        -\frac{m_P^2}{16\pi}R(G_{\mu\nu})
        +\frac12(\bar{\nabla}\phi)^2\right\}
        +\frac{m_P^2}{8\pi}\int_{\partial\cal M}
        d^3x\, ({}^3G)^{1/2} \bar{K}                  \label{3.4}
        \end{eqnarray}
has a minimal coupling and the new inflaton potential
        \begin{eqnarray}
        \bar{V}(\phi)=\left.\left(\frac{m_P^2}{16\pi}\right)^2
        \frac{V(\varphi)}{U^2(\varphi)}
        \,\right|_{\varphi=\varphi(\phi)}.            \label{3.5}
        \end{eqnarray}
The bar indicates here that the corresponding quantity is calculated in the
Einstein frame of fields $(G_{\mu\nu},\phi)$. In what follows we shall
always denote the inflaton field in the Einstein frame (or in the minimal
model) by $\phi$ and that of the nonminimal frame by $\varphi$.

The above transition to the Einstein frame allows one to find the
Hawking-Turok instanton for the model (\ref{3.1}) by transforming the
results of the previous section to the nonminimal frame. Here we shall do it
in the case of a big negative nonminimal coupling $|\xi|\gg 1$ and quartic
potential of (\ref{0.2}):
        \begin{eqnarray}
        &&U(\varphi)=\frac{m_P^2}{16\pi}
        +\frac12 |\xi|\,\varphi^2,                    \label{3.6}\\
        &&V(\varphi)=\frac{m^2\varphi^2}2
        +\frac{\lambda\varphi^4}4
        \end{eqnarray}
The integration of eq.(\ref{3.3}) is rather complicated, but for
large values of the inflaton field, $|\xi|\,\varphi^2/m_P^2\gg 1$, it expresses
$\phi$ in terms of the Eistein frame field $\phi$
        \begin{eqnarray}
        \varphi(\phi)\simeq\frac{m_P}{|\xi|^{1/2}}
        \exp\left[\sqrt{4\pi/3}
        \Big(1+\frac1{6\,|\xi|}\Big)^{-1/2}
        \frac\phi{m_P}\right],                       \label{3.7}
        \end{eqnarray}
where the integration constant is chosen so that the above range of $\phi$
corresponds to $\phi\gg m_P$. The potential in the Einstein frame
equals
        \begin{eqnarray}
        \bar{V}(\phi)=\frac{\lambda m_P^4}{256\pi^2|\xi|^2}
        \,\left[\,1-\frac{1+\delta}{4\pi}\frac{m_P^2}
        {|\xi|\,\varphi^2}+...\right]
        _{\,\phi=\phi(\phi)},                  \label{3.8}
        \end{eqnarray}
where we have retained only the first order term in $m_P^2/|\xi|\,\phi^2$.
In view of (\ref{3.7}), for large $\phi$ this potential exponentially
approaches a constant and satisfies the slow roll approximation with the
expansion parameter
        \begin{eqnarray}
        \epsilon=\frac{m_P}{\sqrt{3\pi}}\frac{\bar V'(\phi_0)}
        {\bar V(\phi_0)}
        \simeq\frac{1+\delta}{3\pi}
        \left(1+\frac1{6\,|\xi|}\right)^{-1/2}
        \frac{m_P^2}{|\xi|\,\varphi_0^2}\ll 1.          \label{3.9}
        \end{eqnarray}

This justifies the above choice of range for the values of the inflaton field.
In this range the Hawking-Turok instanton is described by the equations of the
previous section for the Einstein frame fields $\bar a(\bar\sigma)$ and
inflaton $\phi(\bar\sigma)$. Here $\bar\sigma$ is the coordinate in the
spacetime interval $d\bar s^2=d\bar\sigma^2+\bar a^2(\bar\sigma)\,
d\Omega^2_{(3)}$ of the Einstein frame metric. In view of (\ref{3.2}) these
intervals are related by the equation $d\bar s^2=(16\pi U/m_P^2) ds^2$, so
that the coordinates and scale factors of both frames\footnote
{Note that the relation (\ref{3.2}) holds in one coordinate system covering
the both conformally related spacetimes, while the coordinates $\bar\sigma$
and $\sigma$ are essentially different.}
are related by $d\bar\sigma=\sqrt{16\pi U/m_P^2}\,d\sigma$ and
$\bar a=\sqrt{16\pi U/m_P^2}\,a$. Combining these equations with the asymptotic
behaviour of the Einstein frame fields at $\bar\sigma\rightarrow\bar\sigma_f$
(eqs. (\ref{2.10})-(\ref{2.12}) rewritten for
$\bar a(\bar\sigma),\phi(\bar\sigma)$ with the potential $\bar{V}(\phi)$)
one can easily find the behaviour of
fields in the nonminimal frame. We give it in the limit of large $|\xi|$:
        \begin{eqnarray}
        &&a(\sigma)\simeq\frac4{m_P}
        \left(\frac{m_P}{\phi_0}\right)^{1+2\alpha}
        \left(\frac{1+\delta}{4\pi\lambda}\right)
        ^{1/4+\alpha/2}\,
        \Big[\,m_P(\sigma_f-\sigma)\,\Big]^{1/2-\alpha}, \label{3.10}\\
        &&\varphi(\sigma)\simeq m_P
        \left(\frac{\phi_0}{m_P}\right)^{1/2+3\alpha}
        \left(\frac{1+\delta}{4\pi\lambda}\right)^{1/8-3\alpha/4}\,
        \Big[\,m_P(\sigma_f-\sigma)\,\Big]^{-1/4+3\alpha/2},\\
        &&\alpha\equiv\frac12\,\frac{\sqrt{1+1/6\,|\xi|}-1}
        {1+3\sqrt{1+1/6\,|\xi|}}\simeq\frac1{96\,|\xi|}\ll 1. \label{3.11}
        \end{eqnarray}

In contrast with the minimal coupling we now have the power singularities
for both fields. For large $|\xi|\gg 1$, in particular, they look like
$a\sim (\sigma_f-\sigma)^{1/2}$ and $\varphi\sim (\sigma_f-\sigma)^{-1/4}$.
The inflaton singularity is thus stronger than the logarithmic one in the
minimal case, while that of the scale factor is softer
($1/2-\alpha\geq 1/3$). Note, by the way, that the coefficient of
strongest singularity of the scalar curvature is also suppressed by $1/|\xi$,
$R\sim (1/|\xi|)(\sigma-\sigma_f)^{-2}$. This
property can be qualitatively explained by the fact that the effective
gravitational constant $(m_P^2+8\pi|\xi|\varphi^2)^{-1}$ tends to zero at
the singularity.

The Hawking-Turok action can now be approximately calculated in the Einstein
frame with the aid of eq.(\ref{2.16}), $I_{HT}(\varphi_0)=\bar I_{HT}(\phi_0)$.
Taking into account that
        \begin{eqnarray}
        \frac{3m_P^4}{8\bar V(\phi_0)}\simeq\frac{96\pi^2|\xi|^2}\lambda
        +\frac{24\pi(1+\delta)|\xi|}{\lambda}\frac{m_P^2}{\varphi_0^2}
        +\frac32\frac{(1+2\delta)^2}\lambda
        \left(\frac{m_P^2}{\varphi_0^2}\right)^2+...        \label{3.12}
        \end{eqnarray}
and using the relation
        \begin{eqnarray}
        \sqrt{\frac3{16\pi}}m_P\frac d{d\phi}\simeq\frac12
        \left(1+\frac1{6\,|\xi|}\right)^{-1/2}\varphi\frac d{d\varphi},
        \end{eqnarray}
one finds that the term with the derivative in (\ref{2.16}) almost cancels the
first subleading term in the expansion (\ref{3.12}) and inverts the sign of the
second order term
        \begin{eqnarray}
        &&\left(1+\sqrt{\frac3{16\pi}}m_P
        \frac d{d\phi_0}\right)\,
        \left(-\frac{3m_P^4}{8\bar{V}(\phi_0)}\right)\simeq
        -\frac{96\pi^2|\xi|^2}\lambda
        -\frac{2\pi(1+\delta)}\lambda \frac{m_P^2}{\varphi_0^2}\nonumber\\
        &&\qquad\qquad\qquad\qquad+\frac32\frac{(1+2\delta)^2}\lambda
        \left(\frac{m_P^2}{\varphi_0^2}\right)^2+...\,.    \label{3.13}
        \end{eqnarray}
The terms of different powers in $m_P^2/\varphi^2$ here turn out to be of the
same order of magnitude in $1/|\xi|$.
Later we shall see that at the probability maximum $m_P^2/\varphi^2\gg 1$,
although $m_P^2/|\xi|\varphi^2\sim\epsilon\ll 1$, which means that
the dominant effect comes from the third term of (\ref{3.13}). This term is
however not reliable because it goes beyond the approximation of
eq.(\ref{2.16}) and belongs to the next order of the slow roll expansion,
$O(\epsilon^2)=O(m_P^6/|\xi|^3\varphi^6)$. Thus we have to construct
this expansion to this order inclusive which will be done in the next two sections.

\section{Slow roll expansion}
\hspace{\parindent}
Here we develop the slow roll expansion in the minimal inflaton model subject
to equations of motion (\ref{2.3})-(\ref{2.4}). The corresponding slow roll
corrections to the DeSitter background in (\ref{2.5}) can be obtained by
expanding these equations in perturbations
        \begin{eqnarray}
        \delta a=\frac1{H_0}\delta\tilde a,\,\,
        \delta\phi=\sqrt{\frac3{16\pi}}m_P\delta\tilde\phi.  \label{a5}
        \end{eqnarray}
In the linear order this gives for dimensionless perturbations
$(\delta\tilde a,\delta\tilde\phi)$ the following initial value problem
        \begin{eqnarray}
        &&\left(\frac{d^2}{d\theta^2}+3\cot\theta\frac{d}{d\theta}\right)
        \delta\tilde\phi=\frac32\epsilon+
        \epsilon_1\delta\tilde\phi,                     \label{a6}\\
        &&\frac d{d\theta}\frac{\delta\tilde a}{\cos\theta}=
        -\frac38\epsilon\tan^2\theta\delta\tilde\phi,   \label{a7}\\
        &&\delta\tilde\phi(0)=
        \frac d{d\theta}\delta\tilde\phi(0)=0,\,\,
        \delta\tilde a(0)=0,                                \label{a8}
        \end{eqnarray}
where $\epsilon$ and $\epsilon_1$ are the parameters characterizing the
steepness of the inflaton potential at $\phi_0$
        \begin{eqnarray}
        &&\epsilon=\frac{m_P}{\sqrt{3\pi}}
        \frac{V'(\phi_0)}{V(\phi_0)},              \label{a9}\\
        &&\epsilon_1=\frac{3m_P^2}{8\pi}
        \frac{V''(\phi_0)}{V(\phi_0)}.
        \end{eqnarray}
When the slope of the inflaton potential is not too steep and
$\epsilon\ll 1$ and $\epsilon_1=O(\epsilon)\ll 1$, one can develop
the slow roll expansion in powers of the parameter $\epsilon$ which
determines the rate of change of the potential
$|\dot V/HV|\simeq 3\epsilon^2/8\ll 1$. In the lowest order
approximation the second term on the right hand side of (\ref{a6}) should
be discarded and the solution reads
        \begin{eqnarray}
        &&\delta\tilde\phi(\theta)=\epsilon
        \left(\frac14 \tan^2\frac\theta2
        -\ln\cos\frac\theta2\right)
        +O\left(\epsilon^2\right),              \label{a11}\\
        &&\delta\tilde a(\theta)=
        O\left(\epsilon^2\right).                  \label{a12}
        \end{eqnarray}

The second order approximation is also straightforward. One expands
eqs.(\ref{2.3}) and (\ref{2.4}) to second order in perturbations and
discards the terms $O(\epsilon^3)$ and $O(\epsilon^4)$ respectively.
As a result one retains the second term on the right hand side of eq.(\ref{a6}),
while the equation for $\delta\tilde a$ acquires extra $O(\epsilon^2)$
and $O(\epsilon^3)$ contributions
        \begin{eqnarray}
        \frac d{d\theta}\frac{\delta\tilde a}{\cos\theta}=
        \frac18\tan^2\theta\left[-3\epsilon\delta\tilde\phi+
        \Big(\frac{d}{d\theta}\delta\tilde\phi\Big)^2\right]-
        \frac18\epsilon_1
        \tan^2\theta(\delta\tilde\phi)^2+O(\epsilon^4).   \label{a13}
        \end{eqnarray}

Equations (\ref{a6}) and (\ref{a13}) subject to initial
conditions (\ref{a8}) are not integrable in elementary functions,
for example,
        \begin{eqnarray}
        &&\delta\tilde\phi(\theta)=\epsilon
        \left(\frac14 \tan^2\frac\theta2
        -\ln\cos\frac\theta2\right)
        +\frac16\,\epsilon\epsilon_1
        \left[\,\frac{11}{12}\tan^2\frac\theta2
        +\frac1{3}\ln\cos\frac\theta2\right.\nonumber  \\
        &&\qquad\qquad\qquad\left.
        -\frac{\ln\cos(\theta/2)}{\sin^2(\theta/2)}
        -\frac12+\int_0^{\sin^2(\theta/2)}
        \frac{dy}y\ln(1-y)\,\right]
        +O(\epsilon^3).                                \label{a14}
        \end{eqnarray}
For our purposes, however, it is sufficient to obtain the leading
behaviour in the vicinity of $\theta=\pi$ --
the point where we shall match two perturbation theories. Another
simplification that we use here follows from the potential
$V(\phi)=\bar V(\phi)$ given by eq. (\ref{3.8})
        \begin{eqnarray}
        V(\phi)=\frac{\lambda m_P^4}{256\pi^2|\xi|^2}
        \,\left[\,1-\frac{1+\delta}{4\pi}
        \exp\Big(-\sqrt{\frac{4\pi}3}
        \frac\phi{m_P}\Big)
        +...\right],\,\,\,|\xi|\gg 1,                  \label{3.8a}
        \end{eqnarray}
(in this section we omit the bar in $\bar V(\phi)$, the argument $\phi$ clearly
indicating that the potential belongs to the Einstein frame). For such
a potential the two steepness parameters $\epsilon$ and $\epsilon_1$ above are not
independent
        \begin{eqnarray}
        \epsilon_1=-\frac32\epsilon+O(\epsilon^2).    \label{a15}
        \end{eqnarray}
Thus, putting $\theta=\pi-\varepsilon$ we finally obtain in the vicinity of
the Hawking-Turok singularity, $\varepsilon\ll 1$,
        \begin{eqnarray}
        &&\phi(\pi-\varepsilon)=
        \phi_0+\sqrt{\frac3{16\pi}}m_P\left[\,\epsilon
        \left(\frac1{\varepsilon^2}
        -\ln\frac{\varepsilon}2-\frac16+...\right)\right.\nonumber\\
        &&\qquad\qquad\qquad\qquad\qquad\qquad+\epsilon^2\left.
        \left(-\frac{11}{12\varepsilon^2}
        +\frac16\ln\frac{\varepsilon}2+...\right)
        +O\left(\epsilon^3\right)\right],         \label{a16} \\
        &&H_0\,a(\pi-\varepsilon)=\sin\varepsilon
        +\epsilon^2\left(-\frac1{6\varepsilon^3}+...\right)+
        +\epsilon^3\left(\frac{11}{36\varepsilon^3}+...\right)
        +O\left(\epsilon^4\right),                         \label{a17}
        \end{eqnarray}
where the dots denote terms of higher powers in $\varepsilon$. Perturbation
corrections diverge at $\varepsilon\rightarrow 0$, and an obvious domain of
applicability of this perturbation theory is
        \begin{eqnarray}
        \frac\epsilon{\varepsilon^2}\ll 1.              \label{a19}
        \end{eqnarray}
This is the domain of the agular coordinate $\theta=\pi-\varepsilon$ in which we
will match the slow roll solution with the perturbative solution near the
singularity.

\section{Perturbation expansion for the Hawking-Turok instanton}
\hspace{\parindent}
The perturbation theory in powers of the slow roll parameter $\epsilon$
near the singularity located close to $\theta=\pi$ is qualitatively
different from the slow roll expansion near $\theta=0$. This is the
expansion in powers of the potential in the vicinity of this point, rather
than only in its gradient. To develop such an expansion we first rewrite the
equations of motion as
        \begin{eqnarray}
        &&\frac d{d\sigma}\left(a^3\dot\phi\right)=a^3 V', \label{1}\\
        &&{\dot a}^2=\frac{4\pi}{3m_P^2}(a\dot\phi)^2
        +1-H^2(\phi)a^2,                              \label{2}
        \end{eqnarray}
and then formally integrate the first of equations to convert it to the
integral form
        \begin{eqnarray}
        a\dot\phi=\frac{m_P A^3}{\sqrt{12\pi}}\frac1{a^2}
        +\frac1{a^2}\int_{\sigma_f}^\sigma d\tilde\sigma\,a^3 V'.  \label{3}
        \end{eqnarray}
Here the integration constant was chosen in terms of the constant $A$ in the
asymptotic behaviour (\ref{2.10a})-(\ref{2.10}) at
$\sigma_f,\,\, a(\sigma_f)=0$. With this choice, this asymptotic behaviour is
just the approximate solution of the equations above, corresponding to
discarding the right hand side of (\ref{1}) and the last two terms on the
right hand side of (\ref{2}).

The key point in finding the perturbation corrections to this approximate
solution consists in guessing a special ansatz for the scale factor $a(\sigma)$,
which contains a new small parameter $\bar\epsilon=O(\epsilon)$. The expansion
in this parameter will generate
the perturbative solution with $A$ and $\Phi$ parametrized by $\bar\epsilon$.
Matching the latter with the perturbative solution of the previous
section will allow us to establish a concrete relation between the new
smallness parameter $\bar\epsilon$ and the old parameter of the slow roll
expansion $\epsilon$, (\ref{3.9}), and in this way find $A$ and $\Phi$ as
functions of $\epsilon=\epsilon(\varphi_0)$. This ansatz has the following
form
        \begin{eqnarray}
        a=\frac{\bar\epsilon^{1/2}}{H_0}\bar g(x),\,\,
        x=\frac{\theta-\theta_f}{\bar\epsilon^{1/2}},     \label{4}
        \end{eqnarray}
where $\bar g(x)$ is the some function of the new argument $x$ replacing
the angular coordinate $\theta$ and the parameter $A$ in terms of
$\bar\epsilon$ reads as
        \begin{eqnarray}
        A^3=\frac{3\bar\epsilon}{H_0^2}.                        \label{5}
        \end{eqnarray}
Note that the new argument $x$ in (\ref{4}) is zero at the singularity
and takes negative values on the Hawking-Turok instanton
$0\leq\theta\leq\theta_f$. With this ansatz
$\dot a=d\bar g/dx,\,a\dot\phi=\bar g d\phi/dx$, $H^2(\phi)a^2=O(\bar\epsilon)$
and the second term in the right hand side of (\ref{3}) is also $O(\bar\epsilon)$.
Therefore the equations acquire the form in which all the perturbative
terms containing the potential $V(\phi)=3m_P^2 H^2(\phi)/8\pi$ and its
gradient are explicitly multiplied by $\bar\epsilon$:
        \begin{eqnarray}
        &&\frac{d\phi}{dx}=\sqrt{\frac3{4\pi}}m_P\frac1{\bar g^3}
        +\frac{\bar\epsilon}{\bar g^3 H_0^2}
        \int_0^x\,d\tilde{x}\,\bar g^3 V',                    \label{6}\\
        &&\left(\frac{d\bar g}{dx}\right)^2=1+\frac1{{\bar g}^4}\,
        \left(1+\sqrt{\frac{4\pi}3}\frac{\bar\epsilon}{m_P H_0^2}
        \int_0^x\, d\tilde{x}\,\bar g^3 V'\right)^2
        -\bar\epsilon\,\frac{H^2(\phi)}{H_0^2}\,{\bar g}^2.  \label{7}
        \end{eqnarray}
In the next two subsections we consequitively solve these equations in the
zeroth and first order approximations sufficient for calculating the
Hawking-Turok action up to $\epsilon^2$-order inclusive.

\subsection{Zeroth order approximation}
\hspace{\parindent}
To begin with, note that zeroth order approximation does not exactly
coincide with the asymptotic solution (\ref{2.10a})-(\ref{2.10}), because
the first term on the right hand side of (\ref{7}) is not perturbative in
$\bar\epsilon$, although it becomes negligible at $\sigma\to\sigma_f$ or
$x\to 0$ when $\bar g(x)\to 0$. This term, however, gives a considerable
contribution at the point $\theta=\pi-\varepsilon$, where we match the two
solutions, and therefore it should not be treated perturbatively. Therefore,
the equation for the function $\bar g(x)$, which we denote in the lowest
order approximation by $g(x)$ without the bar, reads
        \begin{eqnarray}
        &&\bar g(x)=g(x)+O(\bar\epsilon), \label{8}\\
        &&\frac{dg(x)}{dx}=-\sqrt{1+\frac1{g^4(x)}}.   \label{9}
        \end{eqnarray}
This function has the asymptotics
        \begin{eqnarray}
        g(x)\simeq (-3x)^{1/3},\,\,x\rightarrow 0,  \label{10}
        \end{eqnarray}
obviously matching with the asymptotic behaviour (\ref{2.10a}). We also
need its behaviour at large negative $x$ corresponding, as we will shortly
see, to the matching point in the domain of validity of the slow roll
expansion (\ref{a19}). It can be obtained by rewriting (\ref{9}) in the
form
        \begin{eqnarray}
        &&g(x)=C-x-\int\limits_{g(x)}^\infty
        dg\left(1-\frac1{1+1/g^4}\right),\\
        &&C\equiv\int_0^\infty dg\,
        \Big(1-\frac1{\sqrt{1+1/g^4}}\Big)=
        \frac{2\pi^{3/2}}{\Gamma^2(1/4)},        \label{12}
        \end{eqnarray}
and solving it for big $g(x)$ by iterations
        \begin{eqnarray}
        &&g(x)=C-x-\frac1{6(C-x)^3}
        -\frac5{168(C-x)^7}+...,\,\,x\rightarrow-\infty.  \label{11}
        \end{eqnarray}
The zeroth order equation for $\phi$
        \begin{eqnarray}
        \frac{d\phi}{dx}=\sqrt{\frac3{4\pi}}m_P\frac1{g^3}  \label{13}
        \end{eqnarray}
cannot be integrated in closed form, because the function $g(x)$ is not
explicitly known, but the scalar field can be found parametrically in
terms of $g(x)$ by integrating the following equation -- the corollary of
(\ref{9}) and (\ref{13})
        \begin{eqnarray}
        \frac{d\phi}{dg}=
        -\sqrt{\frac3{4\pi}}m_P\frac1{g\sqrt{1+g^4}}.      \label{14}
        \end{eqnarray}
Its integration gives the result
        \begin{eqnarray}
        \phi=\Phi+\frac{m_P}{\sqrt{12\pi}}
        \frac12\ln\left(\frac{9H_0^2}{8\bar\epsilon}\right)
        -\sqrt{\frac3{16\pi}}m_P\ln\frac{g^2}{1+\sqrt{1+g^4}},  \label{15}
        \end{eqnarray}
where the integration constant follows from comparing the limit of
$x=(\theta-\theta_f)/\bar\epsilon^{1/2}\rightarrow 0$ with the asymptotic
behaviour (\ref{2.10}).

Matching the solution with its slow roll perturbation expansion in
$\epsilon$, (\ref{a16})-(\ref{a17}), at $\theta=\pi-\varepsilon$ with
$\epsilon/\varepsilon^2\ll 1$ takes place at
        \begin{eqnarray}
        -x\big|_{\theta=\pi-\varepsilon}=
        \frac{\varepsilon-(\pi-\theta_f)}{\bar\epsilon^{1/2}}\gg 1, \label{16}
        \end{eqnarray}
where the expansion (\ref{11}) for $g(x)$ works. Comparison of the scale
factor (\ref{a17}) with (\ref{4})-(\ref{11}),
$H_0 a=(\bar\epsilon^{1/2} C-\pi+\theta_f)+\varepsilon+O(\bar\epsilon^2)$,
shows that the angular coordinate of the Hawking-Turok singularity equals
        \begin{eqnarray}
        \theta_f=\pi-\bar\epsilon^{1/2} C+O(\epsilon).     \label{17}
        \end{eqnarray}
With this value of $\theta_f$,
$g(x)=\varepsilon/\bar\epsilon^{1/2}+O(\bar\epsilon^{1/2})\gg1$. Therefore,
matching the inflaton field (\ref{15}) with (\ref{a16}) ($\varepsilon^0$ and
$\varepsilon^{-2}$ parts) gives $\Phi$ and $\bar\epsilon$ in the lowest
order approximation
        \begin{eqnarray}
        &&\Phi=\phi_0-\frac{m_P}{\sqrt{12\pi}}\frac12
        \ln\left(\frac{9H_0^2}{8\bar\epsilon}\right)+O(\epsilon),\\
        &&\bar\epsilon=\epsilon+O(\epsilon^2).             \label{18}
        \end{eqnarray}
In view of (\ref{5}) these expressions correspond to the relations
(\ref{2.12a})-(\ref{2.12}) already advocated in Sect.2.

\subsection{First order approximation}
\hspace{\parindent}
In the first order approximation the second term of eq.(\ref{6}) can be
obtained by using the zeroth order results. In view of the approximate
expression for the inflaton potential (\ref{3.8a}) one has
        \begin{eqnarray}
        V'(\phi)=\epsilon\,\frac{3^{3/2}m_P}{8\sqrt\pi}H_0^2
        \exp\left[-\sqrt{\frac{16\pi}3}
        \frac{\phi-\phi_0}{m_P}\right]+O(\epsilon^2),   \label{19}
        \end{eqnarray}
where
        \begin{eqnarray}
        \exp\left[-\sqrt{\frac{16\pi}3}
        \frac{\phi-\phi_0}{m_P}\right]=
        \frac{g^2}{1+\sqrt{1+g^4}}+O(\epsilon)     \label{20}
        \end{eqnarray}
due to the results of the zero order approximation. Therefore
        \begin{eqnarray}
        \sqrt{\frac{4\pi}3}\frac{\bar\epsilon}{m_P H_0^2}
        \int_0^x\, d\tilde{x}\,\bar g^3 V'=
        -\frac34\epsilon^2\int_0^{\bar g(x)}\,
        \frac{dg\,g^7}{\sqrt{1+g^4}(1+\sqrt{1+g^4})}
        +O(\epsilon^3),                                  \label{21}
        \end{eqnarray}
where the change of integration variable is approximately done with
the aid of eq.(\ref{9}). Performing the integration we convert the
equation (\ref{6}) to the form
        \begin{eqnarray}
        \frac{d\phi}{dx}=\sqrt{\frac3{4\pi}}m_P\frac1{\bar g^3}
        \,\left[1-\frac3{16}\epsilon^2(\sqrt{1+g^4}-1)^2\right]
        +O(\epsilon^3).                               \label{22}
        \end{eqnarray}
This equation can be used in (\ref{7}) along with the expression
        \begin{eqnarray}
        \frac{H^2(\phi)}{H_0^2}=1+\frac34\,\epsilon\,
        \frac{g^2+1-\sqrt{1+g^4}}{g^2}+O(\epsilon^2)    \label{23}
        \end{eqnarray}
valid in view of the relation (\ref{20}), whence
        \begin{eqnarray}
        \left(\frac{d\bar g}{dx}\right)^2=1+\frac1{{\bar g}^4}
        -\!\bar\epsilon\,\bar g^2\!
        -\frac38\bar\epsilon^2\,
        \left[\,2\,(\bar g^2+1-\sqrt{1+\bar g^4})
        +\frac{(\sqrt{1+\bar g^4}-1)^2}{\bar g^4}\,\right]
        +O(\epsilon^3).                                 \label{24}
        \end{eqnarray}

Integration of (\ref{22}) can again be done parametrically in terms of $\bar g$
with $d\bar g(x)/dx$ taken from (\ref{24}). With the
integration constant again defined in the limit of $\bar g\rightarrow 0$
one has
        \begin{eqnarray}
        &&\phi=\Phi+\frac{m_P}{\sqrt{12\pi}}
        \frac12\ln\left(\frac{9H_0^2}{8\bar\epsilon}\right)
        -\sqrt{\frac3{16\pi}}m_P
        \ln\frac{\bar g^2}{1+\sqrt{1+\bar g^4}}\nonumber\\
        &&\qquad\qquad
        -\bar\epsilon\,\sqrt{\frac3{16\pi}}m_P\frac12 \left[\,
        \ln(\bar g^2+\sqrt{1+\bar g^4})-\frac{\bar g^2}{\sqrt{1+\bar g^4}}\,
        \right] \nonumber\\
        &&\qquad\qquad+\bar\epsilon^2\,\sqrt{\frac3{16\pi}}m_P\frac3{16}
        \left[-2\ln\frac{1+\sqrt{1+\bar g^4}}2
        +\sqrt{1+\bar g^4}-1\,\right]+O(\epsilon^3).      \label{25}
        \end{eqnarray}
The solution of eq.(\ref{24}) with the initial condition $\bar g(0)=0$
can be found in several steps. First it can be integrated perturbatively in
$\bar\epsilon$. For large $\bar g\gg 1$ this integration gives the
equation
        \begin{eqnarray}
        &&\bar g+\frac1{6\bar g^3}-\frac3{56\bar g^7}+...+
        \bar\epsilon\,\left[\frac{\bar g^3}6
        -\frac{5\Gamma^2(1/4)}{48\sqrt\pi}
        +\frac3{4\bar g}+...\right]            \nonumber\\
        &&+\bar\epsilon^2\,\left[\frac3{40}\bar g^5-\frac38 \bar g+
        \left(\frac{63\pi^{3/2}}{40\Gamma^2(1/4)}
        -\frac{5\Gamma^2(1/4)}{64\sqrt\pi}\right)
        +\frac9{16\bar g}+...\right]=C-x+O(\bar\epsilon^3),   \label{26}
        \end{eqnarray}
which, when solved iteratively in $\bar\epsilon$, yields the solution
        \begin{eqnarray}
        &&\bar g(x)=C-x-\frac1{6(C-x)^3}-\frac5{168(C-x)^7}+...\nonumber\\
        &&\qquad\qquad\qquad+\bar\epsilon\,\left[-\frac16 (C-x)^3
        +\frac{5\pi}{24C}+O(1/(C-x)^5)\right]\nonumber\\
        &&\qquad\qquad\qquad+\bar\epsilon^2\,\left[\frac1{120}(C-x)^5-
        \frac{5\pi}{48C}(C-x)^2+O(C-x)\right]+O(\bar\epsilon^3)   \label{27}
        \end{eqnarray}
expanded in inverse powers of $(C-x)$. Matching
this solution with the slow roll perturbation expansion
(\ref{a16})-(\ref{a17}) at $\theta=\pi-\varepsilon$ begins by noting that
at the matching point
        \begin{eqnarray}
        &&C-x=\frac{\varepsilon-\varepsilon_0}
        {\bar\epsilon^{1/2}},                        \label{28}\\
        &&\varepsilon_0\equiv\pi
        -\theta_f-C\bar\epsilon^{1/2}.              \label{29}
        \end{eqnarray}
From the zeroth order approximation we know that
$\varepsilon_0=O(\bar\epsilon)$, therefore the scale factor (\ref{4})
with (\ref{11}) at this point takes the form of double series in
$\bar\epsilon$ and $\varepsilon$.
        \begin{eqnarray}
        &&H_0 a(\pi-\varepsilon)=\varepsilon
        -\frac16\varepsilon^3+\frac1{120}
        \varepsilon^5+...-\varepsilon_0\,
        \left(1-\frac36\varepsilon^2+...\right)\nonumber\\
        &&\qquad\qquad+\bar\epsilon^{3/2}\left(\frac{5\pi}{24C}
        -\frac{5\pi}{48C}\varepsilon^2+...\right)+\bar\epsilon^2
        \left(-\frac1{6\varepsilon^3}+...\right)+
        O\left(\bar\epsilon^{5/2}\right)              \label{30}
        \end{eqnarray}
Comparison with (\ref{a17}) then shows that the first group of terms
reproduces its unperturbed part, $\sin\varepsilon$, while the choice of
        \begin{eqnarray}
        \varepsilon_0=\frac{5\pi}{24C}\,
        \bar\epsilon^{3/2}+O\left(\bar\epsilon^2\right)    \label{31}
        \end{eqnarray}
guarantees the absence of the half integer power of
$\bar\epsilon$. Eq.(\ref{30}) cannot serve, however, for the determination
of $\bar\epsilon$ in terms of $\epsilon$, because it lacks the next
-- $\bar\epsilon^3$ -- order of perturbation theory. But the corresponding
-- $\bar\epsilon^2$ -- order is already contained in the expression
(\ref{25}) for $\phi$. Therefore, its matching with (\ref{a16}) can be
used for the determination of both unknown quantities $\Phi$ and
$\bar\epsilon$.

With the above choice of parameters we have
$\bar g=\sin\varepsilon/\bar\epsilon^{1/2}+O(\bar\epsilon^{3/2})$ at the
matching point and the inflaton (\ref{25}) takes the form
        \begin{eqnarray}
        &&\phi(\pi-\varepsilon)=\Phi+\frac{m_P}{\sqrt{12\pi}}
        \frac12\ln\left(\frac{9H_0^2}{8\bar\epsilon}\right)\nonumber\\
        &&\qquad
        +\sqrt{\frac3{16\pi}}m_P\left[\,\bar\epsilon
        \left(\frac1{\varepsilon^2}
        +\frac12\ln\frac{\bar\epsilon}{2\varepsilon^2}
        +\frac12+...\right)
        +\bar\epsilon^2\left(\frac38\ln\frac{2\bar\epsilon}
        {\varepsilon^2}-\frac3{16}+...\right)
        +O\left(\bar\epsilon^3\right)\right].            \label{33}
        \end{eqnarray}
Now we match this expression with (\ref{16}). The both double pole and
logarithmic in $\varepsilon$ terms here dutifully coincide with those of
(\ref{a16}) under the following identification
        \begin{eqnarray}
        \bar\epsilon=\epsilon-\frac{11}{12}\,\epsilon^2
        +O\left(\epsilon^3\right)                          \label{32}
        \end{eqnarray}
and $\Phi$ takes the form
        \begin{eqnarray}
        \Phi(\phi_0)=\phi_0-\frac12\frac{m_P}{\sqrt{12\pi}}
        \ln\left[\frac{9H_0^2}{8\epsilon}\right]
        -\frac12\sqrt{\frac3{16\pi}}m_P\,\epsilon
        \left(\ln\frac{\epsilon}8
        +\frac{35}{18}\right)+O\left(\epsilon^2\right).    \label{3.15}
        \end{eqnarray}
In view of (\ref{5}) and (\ref{31}) the other two parameters of the asymptotic
behaviour near the singularity read
        \begin{eqnarray}
        &&\theta_f\equiv H_0\sigma_f\simeq\pi
        -\frac{2\pi^{3/2}}{\Gamma^2(1/4)}
        \epsilon^{1/2}-\frac{5\Gamma^2(1/4)}{48\pi^{1/2}}
        \epsilon^{3/2},                                \label{3.14}\\
        &&A^3(\phi_0)\simeq\frac{3\epsilon}{H_0^2}
        -\frac{11}4\frac{\epsilon^2}{H_0^2}.            \label{3.15a}
        \end{eqnarray}

The expressions (\ref{3.15}) and (\ref{3.15a}) can now be substituted into
the equation (\ref{2.15a}) for $\bar I_{HT}(\phi_0)$. Then the
integration of terms belonging to the zeroth order approximation will
obviously reproduce the contribution (\ref{3.13}) which is already
written in terms of the inflaton field in the nonminimal frame. The first
order corrections will generate the contribution
$\Delta\bar I_{HT}(\phi_0)=\Delta I_{HT}(\varphi_0)$ which, unfortunately,
can no longer be obtained in a closed form for an arbitrary potential.
Rather, the potential (\ref{3.8a}) has to be used in (\ref{2.15a}).
The integration and the conversion of the result to the original frame then
gives the final result for the Hawking-Turok action to the second order
in $m_P^2/|\xi|\varphi^2$ inclusive
        \begin{eqnarray}
        &&I_{HT}(\varphi)=-\frac{96\pi^2|\xi|^2}\lambda
        -\frac{2\pi(1+\delta)}\lambda \frac{m_P^2}{\varphi^2}
        +2\frac{(1+\delta)^2}\lambda
        \left(\frac{m_P^2}{\varphi^2}\right)^2
        \ln\left(\frac{6\pi|\xi|\varphi^2}
        {m_P^2(1+\delta)\kappa}\right)\nonumber\\
        &&\qquad\qquad\qquad\qquad\qquad\qquad\qquad\qquad
        +O\,\left(\frac{m_P^6}{|\xi|\varphi^6}\right),  \label{3.16}
        \end{eqnarray}
where $\kappa$ absorbs the following combination of numerical parameters
        \begin{eqnarray}
        \ln\kappa\equiv\frac{11}3-\ln 4
        -\frac34\left(\frac{1+2\delta}{1+\delta}\right)^2.
        \end{eqnarray}
Due to big $|\xi|$ it contains a large but slowly varying (in $\varphi$)
logarithmic coefficient. The positive coefficient of this logarithmic
term actually follows from the sign of $\ln\cos(\theta/2)$ in the
equation (\ref{a11}) for $\delta\phi$ above and, thus, it is pretty well
fixed. This sign will have important consequences for quantum
creation of the open Universe.

\section{Tree-level probability peaks}
\hspace{\parindent}
Note, that the logarithmic structure of the result (\ref{3.16}) resembles the
behaviour of Coleman-Weinberg loop effective potentials (up to inversion of
$\varphi$), even though this
term is entirely of a tree-level origin. Thus, the classical theory somehow
feels quantum structures when probing Planckian scales near the singularity.
The nontrivial structure of the classical action allows one to expect the
existence of the probability peaks already in the tree-level approximation
with the distribution functions
$\rho_{NB,T}(\varphi)\simeq\exp(\mp I_{HT}(\varphi))$. The extremum
equation for the exponential
        \begin{eqnarray}
        &&\frac d{dx}I_{HT}(\varphi(x))=
        \frac{12\pi^2|\xi|}{\lambda\kappa}\frac1{x^3}
        \left(x-\frac{12|\xi|}\kappa\ln\frac x{\sqrt e}\right)=0, \label{7.1}\\
        &&x\equiv\frac{6\pi|\xi|\varphi^2}
        {m_P^2(1+\delta)\kappa},                \label{7.2}
        \end{eqnarray}
indeed results in two roots $x_\mp$ which for big $|\xi|$ read as
        \begin{eqnarray}
        &&x_-\simeq\sqrt e,              \label{7.3}\\
        &&x_+\simeq\frac{12|\xi|}{\kappa}
        \ln\left(\frac{12|\xi|}{\kappa\sqrt e}\right).   \label{7.4}
        \end{eqnarray}

To analyze the inflationary scenario generated by these peaks we have to
resort to the classical equations of motion. For the model (\ref{3.1}) they were
considered in much detail in \cite{efeq}. The slope of the
potential (\ref{3.8}) is positive for $\delta>-1$ which implies the
finite inflationary epoch with slowly decreasing inflaton
only in this range of $\delta$. The inflationary e-folding number in this case
is approximately given by the equation
        \begin{eqnarray}
        N\simeq\int_0^{\varphi_I}d\varphi\,
        \frac{3H^2(\varphi)}{|F(\varphi)|},             \label{5.4}
        \end{eqnarray}
where the Hubble constant
        \begin{eqnarray}
        H^2(\varphi)\simeq \frac\lambda{12|\xi|}\,\varphi^2
        \end{eqnarray}
and $F(\varphi)$ is the rolling force in the inflaton equation of motion
for the nonminimal model $\ddot\varphi+3H\dot\varphi-F(\varphi)=0$,
$F(\varphi)\simeq -\lambda m_P^2(1+\delta)\varphi/48\pi\xi^2$.
The integration then gives
        \begin{eqnarray}
        N\simeq\frac{6\pi|\xi|\varphi^2}
        {m_P^2(1+\delta)},                           \label{5.4a}
        \end{eqnarray}
so that the parameter $x$ introduced for brevity in (\ref{7.2}) is related
to the e-folding number, $N=\kappa x$. The order of magnitude of the
coupling constants, we shall use, is the one that fits the COBE
normalization for the microwave background anisotropy generated in this
model and good conditions for reheating \cite{nonmin}, $|\xi|\sim 2\times 10^4$,
$\lambda\sim 0.05$, $\sqrt\lambda/|\xi|\sim 10^{-5}$. We shall also assume
that the mass term in the inflaton potential is small, so that the
parameter $\delta$ in (\ref{delta}) is negligible in what follows. With
this choice the parameter $\kappa$ is approximately given by
        \begin{eqnarray}
        \kappa\simeq 4.6.
        \end{eqnarray}

For the smaller root $x_-$, the second derivative of the action is
negative, $d^2I_{HT}/dx^2=-(12\pi|\xi|/\kappa e)^2/\lambda<0$, so that
it corresponds to the maximum of the {\em tunneling} distribution
$\rho_{T}\sim\exp(+I_{HT})$. This tunneling peak has a a very small relative width
$\Delta x_-/x_-\simeq \kappa\sqrt{\lambda e}/12\pi|\xi|\sim 10^{-6}$ and
generates a very small e-folding number for inflationary stage beginning
with the GUT scale of the Hubble constant
        \begin{eqnarray}
        &&N_-=\kappa x_-\simeq\kappa\sqrt e\sim 9,\\
        &&H^2_-\simeq m_P^2\frac\lambda{|\xi|^2}
        \frac{N_-(1+\delta)}{72\pi}\sim 10^{-11}m_P^2.
        \end{eqnarray}
According to (\ref{i1}) such an e-folding number generates extremely small
value of $\Omega$, $\Omega\sim e^{-110}$, which makles this peak
unacceptable as a candidate for the initial conditions for open inflation.

The second root (\ref{7.4}) brings us to another extreme. Because of the
positive second derivative
$d^2I_{HT}(x_+)/dx_+^2=\pi^2\kappa^2/144
\lambda|\xi|^2[\ln(12|\xi|/\kappa \sqrt e)]^3$ this peak exists for the
no-boundary case with $\rho_{NB}\sim\exp(-I_{HT})$. This peak is rather
smeared, for its relative width is $\Delta x_+/x_+\sim\sqrt\lambda\sim 0.2$,
and its parameters are
        \begin{eqnarray}
        &&N_+=12|\xi|\ln\frac{12|\xi|}{\kappa\sqrt e}\sim 10^6,\\
        &&H^2_+\simeq m_P^2\frac\lambda{|\xi|^2}
        \frac{N_-(1+\delta)}{72\pi}\sim 10^{-11}m_P^2.
        \end{eqnarray}
The energy scale again belongs to the GUT domain, but the corresponding
density parameter is far too close to unity,
$\Omega\sim 1-e^{-2\times 10^6}$, to account for the observable Universe.

\section{Quantum corrections}
\hspace{\parindent}
The most vulnerable point of the Hawking-Turok instanton is the construction
of quantum corrections on its singular background. Although the classical
Euclidean action is finite, the quantum part of the effective action involving
the higher order curvature invariants is infinite, because their spacetime
integrals are not convergent at the singularity. At least
naively, this means that the whole amplitude is either suppressed to zero or
infinitely diverges indicating strong instability. Clearly, a self-consistent
treatment should regularize the arising infinities due
to the back reaction of the infinitely growing quantum stress tensor.

The result of such a self-consistent treatment is hardly predictable because
we do not yet have for it an exhaustive theoretical framework. This
framework might include fundamental stringy structures underlying our local
field theory, which are probed by Planckian curvatures near the singularity.
However, even without the knowledge of this fundamental framework it is worth
considering usual quantum corrections due to local fields
on a given singular background. This might help revealing those dominant
mechanisms that are robust against the presence of singularities and their
regulation due to back reaction and fundamental strings.

These quantum corrections can be devided into two main categories -- nonlocal
contributions due to massless or light degrees of freedom and local
contributions due to heavy massive fields \cite{beyond}. The effects from the first
category can be exactly calculable when they are due to the conformal anomaly
of the conformal invariant fields. For the Hawking-Turok instanton this
calculation can be based on the (singular) conformal transformation mapping
its geometry to the regular metric $d\tilde{s}^2$ of the half-tube
$R^+\times S^3$,
        \begin{eqnarray}
        ds^2=a^2(\sigma(X))\,d\tilde{s}^2,\,\,\,\,
        d\tilde{s}^2=dX^2+d\Omega_{(3)}^2,               \label{4.1}
        \end{eqnarray}
with the conformal coordinate $X=\int_\sigma^{\sigma_f}d\sigma'/a(\sigma'),\,
0\leq X<\infty$. With this conformal decomposition of the metric the effective
action $\SGamma[\,g_{\mu\nu}]$ can be represented as a sum of the finite
effective action of $\tilde{g}_{\mu\nu}$, $\SGamma[\,\tilde{g}_{\mu\nu}]$,
and $\Delta\SGamma[\,\tilde{g}_{\mu\nu},a]$ -- the anomalous action obtained
by integrating the known conformal anomaly along the orbit of the local
conformal group joining $g_{\mu\nu}$ and $\tilde{g}_{\mu\nu}$. The anomalous
action $\Delta\SGamma[\,\tilde{g}_{\mu\nu},a]$ is known for problems without
boundaries \cite{Riegert,Tsconf,BMZ}. For a singular conformal factor
$a^2(\sigma(X))$ the bulk part of this action is divergent, but the question
of its finiteness is still open, because in problems with
boundaries the conformal anomaly has surface (simple and double
layer) contributions \cite{Dowker} that might lead to finite anomalous
action on the Hawking-Turok instanton \cite{Bconf}.

Fortunately, the problem with large nonminimal coupling of the inflaton
falls into the second category of problems -- local effective action of
heavy massive fields. Due to the Higgs mechanism for all matter fields
interacting with inflaton by (\ref{0.3}) their particles acquire masses
$m^2\sim\varphi^2$ strongly exceeding the spacetime curvature
$R\sim\lambda\varphi^2/|\xi|$ \cite{qsi,qcr,efeq}. The renormalized
effective action expanded in powers of the curvature to mass squared ratio
$R/m^2\sim\lambda/|\xi|\ll 1$ for generic spacetime background has the
following form of the local Schwinger-DeWitt expansion \cite{DW,BarvV,efeq}
        \begin{eqnarray}
        &&\mbox{\boldmath$\SGamma$}^{\rm 1-loop}=
        \frac1{32\pi^2}\,\int d^4x\,g^{1/2}\,
        {\rm tr}\,\left\{\frac12\left(\ln\frac{m^2}{\mu^2}-\frac32
        \right)\,m^4\hat{1}\right.\nonumber\\
        &&\quad\qquad+\left(\ln\frac{m^2}{\mu^2}-1
        \right)\,m^2\hat{a}_1(x,x)
        +\ln\frac{m^2}{\mu^2}\,
        \hat{a}_2(x,x)
        -\left.\sum_{n=1}^{\infty}\frac{(n-1)!}{m^{2n}}\,
        \hat{a}_{n+2}(x,x)\,\right\}.                     \label{4.2}
        \end{eqnarray}
Here ${\rm tr}$ denotes the trace over isotopic field indices, hats denote the
corresponding matrix structures in vector space of quantum fields and
$\hat{a}_n(x,x)$ are the Schwinger-DeWitt coefficients. The latter can be
systematically calculated for generic theory as spacetime invariants of
growing power in spacetime and fibre bundle curvatures
\cite{DW,Gilkey,BarvV,Avram}.

The situation with this expansion on a singular
instanton of Hawking and Turok is also not satisfactory -- all integrals of
curvature invariants starting with $\hat{a}_2(x,x)$ (quadratic in the
curvature and higher) diverge at the singularity. Fortunately, the lessons
from the asymptotic theory of semiclassical expansion teach us that the
lowest order terms can be trusted as long as they are well defined. In our
case this is the term quartic in masses with the logarithm. But this term
is exactly responsible for the dominant contribution to the anomalous
scaling (\ref{1.1}) quadratic in $|\xi|$,
        \begin{eqnarray}
        Z=\frac1{32\pi^2}\int d^4x\,g^{1/2} \sum_{\rm particles}
        m^4+...\,.
        \end{eqnarray}
This term is dominating the quantum
part of effective action, while the others, although being divergent at the
singularity, are strongly suppressed by powers of $1/|\xi|$. With the
assumption that the back reaction of quantum stress tensor regulates these
divergences, one can conclude that the quantum effective action on the
Hawking-Turok instanton in our model is still dominated by the anomalous
scaling term of eq.(\ref{1.1}). In the next section we analyze the
consequences of this result.

\section{Open inflation without anthropic principle}
\hspace{\parindent}
Consider now the distribution functions (\ref{1.1}) with the Hawking-Turok
action (\ref{3.16}) and the anomalous scaling term of
$\mbox{\boldmath$\SGamma$}^{\rm 1-loop}$.
A crucial difference from the case of closed cosmology is that the
$\varphi$-dependence of $I_{HT}(\varphi)$ is dominated now by
$|\xi|^0$-term which contains due to a big slowly varying logarithm
a large {\it positive} contribution quartic in $m_P/\varphi$. In the case
of the no-boundary distribution function the extremum equation reduces to
        \begin{eqnarray}
        \frac d{dx}(I_{HT}+\mbox{\boldmath$\SGamma$}^{\rm 1-loop})=
        \frac{3|\xi|^2\mbox{\boldmath$A$}}{\lambda\, x^3}
        \left(x^2-\frac{24\pi^2}{\kappa^2 e\,\mbox{\boldmath$A$}}
        \ln\frac{x^2}e
        +\frac{4\pi^2 x}{\kappa|\xi|\mbox{\boldmath$A$}}\right)=0  \label{9.1}
        \end{eqnarray}
with the same variable $x$ replacing $\varphi$ as in (\ref{7.2}). For large
$|\xi|\gg 1$ the last term can be omitted, and the equation again has two
roots which equal, in the assumption that
$24\pi^2/\kappa^2 e\,\mbox{\boldmath$A$}\gg 1$,
        \begin{eqnarray}
        &&x^2_-\simeq e,              \label{9.2}\\
        &&x^2_+\simeq
        \frac{24\pi^2}{\kappa^2 e\,\mbox{\boldmath$A$}}
        \ln\frac{24\pi^2}{\kappa^2 e\,\mbox{\boldmath$A$}}.   \label{9.3}
        \end{eqnarray}
The first root $x_-$ does not yield the probability maximum because
$d^2(I_{HT}+\mbox{\boldmath$\SGamma$})/dx^2
\sim-144\pi^2|\xi|^2/\kappa^2\lambda<0$, while the second root $x_+$
generates the peak, for
        \begin{eqnarray}
        \left.\frac{d^2}{dx^2}(I_{HT}+\mbox{\boldmath$\SGamma$})
        \right|_{x_+}\simeq
        \frac{144\pi^2|\xi|^2}{\kappa^2\lambda}
        \left(\ln\frac{24\pi^2}
        {\kappa^2 e\,\mbox{\boldmath$A$}}-1\right)>0.   \label{9.4}
        \end{eqnarray}
The positivity of this expression and the asumption above will soon be justified
from the bound on the e-folding number. The parameters of this peak
 -- the mean value $\varphi_I$, Hubble constant $H_I$ and quantum dispersion
$\Delta\varphi\equiv[d^2(I_{HT}+\mbox{\boldmath$\SGamma$})/d\varphi_I^2]^{-1/2}$
-- are
        \begin{eqnarray}
        &&\varphi_I^2\simeq \frac{m_P^2}{|\xi|}\,
        \left(\frac2{3\mbox{\boldmath$A$}}
        \ln\frac{24\pi^2}
        {\kappa^2 e\,\mbox{\boldmath$A$}}
        \right)^{1/2},                                   \label{5.2}\\
        &&H_I^2\simeq m^2_P\,\frac\lambda{|\xi|^2}\,
        \frac1{12}
        \left(\frac2{3\mbox{\boldmath$A$}}
        \ln\frac{24\pi^2}
        {\kappa^2 e\,\mbox{\boldmath$A$}}\right)^{1/2},              \\
        &&\frac{\Delta\varphi}{\varphi_I}
        \sim\frac{\Delta H}{H_I}\sim
        \frac{\kappa^2\sqrt{6\mbox{\boldmath$A$}}}{72\pi^2}
        \,\frac{\sqrt\lambda}{|\xi|}.                      \label{5.3}
        \end{eqnarray}
Similarly to the closed model, these parameters are suppressed relative to
the Planck scale by a small dimensionless ratio $\sqrt{\lambda}/|\xi|$ known
from the COBE normalization. As regards the e-folding number (\ref{5.4a})
it is given for this peak entirely in terms of the same universal combination
of coupling constants (\ref{A}) (remember that, apart the negligible
dependence of $\kappa$ on $\delta$, $\kappa\sim 4.6$)
        \begin{eqnarray}
        N\simeq\left(\frac{24\pi^2}{\mbox{\boldmath$A$}}
        \ln\frac{24\pi^2}
        {\kappa^2 e\,\mbox{\boldmath$A$}}\right)^{1/2}.   \label{5.5}
        \end{eqnarray}
Comparison of this result with the e-folding number, $N\sim 60$, necessary
for generating the observable density $\Omega$, $0<\Omega<1$, not very close
to one or zero, immeadiately gives the bound on $\mbox{\boldmath$A$}$
        \begin{eqnarray}
        \mbox{\boldmath$A$}\sim\frac{48\pi^2}{N^2}
        \ln\frac{N}{\kappa e}\sim 0.3.                         \label{5.6}
        \end{eqnarray}
This bound justifies the above assumption on the magnitude of the factor
$24\pi^2/\kappa^2 e\,\mbox{\boldmath$A$}\sim 10\gg 1$, the positivity in (\ref{9.4})
and the validity of the slow roll approximation throughout the whole paper -- the
expansion in powers of $m_P^2/4\pi|\xi|\varphi_I^2\sim \ln N/N$ (see
eqs.(\ref{3.8}) and (\ref{3.13})).

A similar analysis for the case of the tunneling distribution function
shows that the corresponding extremum equation has the same form (\ref{9.1})
but with opposite signs of the last two terms inside the brackets.
Therefore, it has only one root $x\simeq\sqrt e$. This is again the
maximum of the distribution function, because
$d^2(-I_{HT}+\mbox{\boldmath$\SGamma$})/dx^2
\sim144\pi^2|\xi|^2/\kappa^2\lambda>0$, but the corresponding e-folding
number, $N\simeq\kappa\sqrt e\sim 8$, and the density parameter
$\Omega\sim e^{-100}$
are far too small to describe the observable Universe. This leaves us with
the only candidate for the initial conditions of inflation (\ref{5.2})-(\ref{5.3})
generated by the no-boundary wavefunction of the open Universe.

\section{Conclusions}
\hspace{\parindent}
Thus, we have derived the no-boundary and tunneling distribution functions
for the open inflationary models originating by the Hawking-Turok
mechanism from the singular instanton. By applying the slow roll perturbation
expansion we calculated the classical action of this instanton in the
model with a strong nonminimal coupling. The resulting distribution
functions have in the tree-level approximation sharp probability peaks,
which are, however, incompatible with the observational bounds on
$\Omega$. The inclusion of one-loop corrections allows one to get such a
probability peak satisfying these bounds for the no-boundary quantum state
of the open Universe. The corresponding probability maximum for the
tunneling state is far out of the observational range because it predicts
absolutely negligible value of $\Omega$.

These above conclusions are based on classical equations of inflationary dynamics.
The latter should certainly be replaced by effective equations for mean fields
to have reliable answers within the same accuracy as the calculation of the
one-loop distribution function. Since quantum
effects qualitatively change the tree-level initial conditions, one should
expect that they might strongly influence the dynamics as well. Effective
equations of motion for the model (\ref{3.1}) were obtained in our recent
paper \cite{efeq}, but according to the discussion of the previous section
they are not, strictly speaking, applicable here. This is because the
Hawking-Turok instanton background does not satisfy the condition of the
local Schwinger-DeWitt expansion\footnote
{For closed cosmology with no-boundary or tunneling quantum states the
slow roll approximation guarantees the validity of the local Schwinger-DeWitt
expansion that was used in \cite{efeq} for the derivation of the effective
equations.
}.
It does not satisfy this condition globally, in the vicinity of the
singularity, but the open inflationary Universe lies inside the light cone
originating from the instanton pole antipoidal to the
singularity. Therefore, if we restrict ourselves with the local part
of effective equations, then for this part we can use our old
results \cite{efeq}. The influence of the spatially remote singularity
domain is mediated by nonlocal terms. These terms strongly depend on the
boundary conditions at the singular boundary and are beyond the control of
the local Schwinger-DeWitt approximation.
Within the same reservations as those concerning the finiteness of quantum
effects on this instanton (and the validity of the Hawking-Turok model as a
whole) we can neglect nonlocal terms and use only the local part of effective
equations.

There are two arguments in favour of this approximation. Firstly,
it is very likely that the nonlocal contribution of the singularity is
suppressed by inverse powers of $|\xi|\gg 1$. At least naively, these effects
are inverse to the size of the Universe given by the Hubble constant in
(\ref{5.3}) and also can be damped by $1/|\xi|$ in the same way as for the
singular part of the scalar curvature. Secondly, if we restrict
ourselves with a limited spatial domain of the very early open Universe
(close to the tip of the light cone originating from the regular pole of the
Hawking-Turok instanton), then these nonlocal terms do not contribute at all
in view of causality of effective equations, because at early moments of time
this domain is causally disconnected from singularity\footnote
{This argument is rigorous but somewhat macabre for the inhabitants of this
domain, because after a while they will suffer a fatal influence of fields
propagating from the singularity.
}.

Local quantum corrections in effective equations depend only on the local geometry
of the quasi-DeSitter open Universe. They boil down to the replacement of
the classical coefficient functions ($V(\varphi),\,U(\varphi)$) of the model
(\ref{3.1}) by the effective ones calculated in \cite{efeq} for a wide class
of quantum fields coupled to the inflaton in the limit of big $|\xi|\gg 1$.
It remains to use these functions in classical equations and study the
inflation dynamics starting from the initial value of the inflaton
(\ref{5.2}). It turns out that unlike in the closed model (where the quantum
terms were of the same order of magnitude as classical ones) the quantum
corrections are strongly suppressed by the slow-roll parameter
$\sqrt{\mbox{\boldmath$A$}}/4\pi\sim \ln N/N$ already at the start of inflation.
In particular, the effective rolling force differs from the classical one
above by the negligible correction
        \begin{eqnarray}
        F_{\rm eff}(\varphi)\simeq
        -\frac{\lambda m_P^2(1+\delta)}{48\pi\xi^2}\,\varphi\,
        \left\{1+\left[\frac{\mbox{\boldmath$A$}}{96\pi^2}
        \ln\frac{24\pi^2}{\kappa^2e\mbox{\boldmath$A$}}
        \right]^{1/2}\frac{\varphi^2}{\varphi_I^2}\right\}.    \label{5.7}
        \end{eqnarray}
For comparison, in the closed model the second (quantum) term in curly
brackets enters with a unit coefficient of $\varphi^2/\varphi_I^2$, see
eqs.(6.7) and (6.10) of \cite{efeq}. This quantum correction gives
inessential contribution to the duration of the inflationary stage
(\ref{5.5}) and thus does not qualitatively change the above predictions
and bounds. The smallness of quantum corrections (roughly proportional to
$\mbox{\boldmath$A$}/32\pi^2$) can be explained by stronger bound on
$\mbox{\boldmath$A$}\sim 0.3$ (cf. $\mbox{\boldmath$A$}\leq 5.5$ in the
closed model \cite{efeq}) and another dependence of $\varphi_I$ on
$\mbox{\boldmath$A$}$.

Thus, the no-boundary quantum state on the Hawking-Turok instanton in the
model with large nonminimal curvature coupling generates open inflationary
scenario compatible with observations and, in particular, capable of
producing
the needed value of $\Omega$. No anthropic considerations or fine tuning of
initial conditions is necessary to reach such a final state of the Universe.
The only fine tuning we get is the bound on the parameter
$\mbox{\boldmath$A$}$ of the matter field sector (\ref{5.6}) and the
estimate of the ratio $\sqrt{\lambda}/|\xi|\sim 10^{-5}$ based on the
normalization from COBE which looks as a natural determination of coupling
constants of Nature from the experiment. The mechanism of such quantum birth
of the Universe is based on quantum effects on the Hawking-Turok instanton,
treated within semiclassical loop expansion. The validity of this expansion
is in its turn justified by the energy scale of the phenomenon (\ref{5.3})
which belongs to the GUT domain rather than the Planckian one.

\section*{Acknowledgements}
\hspace{\parindent}
The author is grateful to A.A.Starobinsky for
useful discussions. Helpful correspondence with A.Linde is also
deeply acknowledged. This work was supported
by the Russian Foundation for Basic Research under grants No 96-02-16287 and
No 96-02-16295, the European Community Grant INTAS-93-493-ext and
by the Russian Research program ``Cosmomicrophysics''.

\end{document}